\documentclass[aps,onecolumn,showpacs,preprintnumbers,floats,amsmath]{revtex4}

\usepackage{amssymb}  % \gtrsim, \geqslant, etc: see amsguide.ps
\usepackage{graphicx}
\usepackage{dcolumn}% Align table columns on decimal point
\usepackage{bm,CJK}
\usepackage{epsfig}
\usepackage{hyperref}
\usepackage{graphicx}% Include figure files
\usepackage{dcolumn}% Align table columns on decimal point
\usepackage{bm}% bold math
\usepackage{longtable}
\usepackage[dvips]{color}

\usepackage[normalem]{ulem}  % \sout{old text} for strikeout

\newcommand{\NPA}[3]{Nucl.\ Phys.\ {\bf A#1}, #2 (#3)}

\newcommand{\PLB}[3]{Phys.\ Lett.\ B\ {\bf #1}, #2 (#3)}

\newcommand{\PRL}[3]{Phys.\ Rev.\ Lett.\ {\bf #1}, #2 (#3)}

\newcommand{\PRC}[3]{Phys.\ Rev.\ C\ {\bf #1}, #2 (#3)}
\newcommand{\PRD}[3]{ Phys.\ Rev.\ D\ {\bf #1}, #2 (#3)}
\newcommand{\JPG}[3]{J.\ Phys.\ G\ {\bf #1}, #2 (#3)}

\begin{document}

\title{ The compressibility of quark matter under strong magnetic field in the NJL model}
\date{\today}

\author{Li Yang$^{1}$ \footnote{yangl@jzxy.edu.cn},
Xin-Jian Wen$^{2}$ \footnote{wenxj@sxu.edu.cn} }

\affiliation{ $^{1}$Department of Physics and Electronic
Engineering,Jinzhong University, Jinzhong 030619, China\\}

\affiliation{$^{2}$ Institute of Theoretical Physics, State Key
Laboratory of Quantum Optics and Quantum Optics Devices,Shanxi
University, Taiyuan 030006, China}

\begin{abstract}
The compressibility of magnetized quark matter is investigated in
the SU(2) NJL model. The increases of the chemical potential and the
temperature can reduce the compressibility, and lead to the much
stiffer equation of state. The variation of the compressibility with
the magnetic field will depend on the phase region. Due to the
anisotropic structure, the compressibility is different in the
directions parallel and perpendicular to the field. The
discontinuity of longitudinal compressibility with the chemical
potential and the temperature captures the signature of a
first-order chiral phase transition and the crossover at high
temperature. Moreover, the magnetic-field-and-temperature running
coupling would have an important effect on the position of the phase
transition. Under the lowest landau level approximation at zero
temperature, the longitudinal compressibility has a direct inverse
proportional relation to the magnetic field strength and the
chemical potential square as
$\kappa^\parallel_{\mathrm{LLL},\chi}\propto1/(eB \mu^2)$.

\end{abstract}

 \maketitle


\section{INTRODUCTION}
The properties of the quark matter are of most importance in
understanding many physical aspects of nature, such as the quark
gluon plasma in the big bang of the early universe, the possible
structure in the core of compact objects, and the hadronic quark
phase transition in experiments, where the high temperature and high
densities characterize the extreme conditions. Recently, the study
of properties of the quark matter is extended to a strong magnetic
field \cite{Miransky15,catapaper,Felipe09}, and further extended to
a parallel electric and magnetic field background \cite{rugg16}. It
is widely accepted that the strong magnetic fields could exist in
the early universe, in the core of neutron stars, and in the
noncentral heavy ion collision experiments, such as the Relativistic
Heavy Ion Collider or the Large Hadron Collider (LHC)
\cite{magnetic}. The magnitude of the magnetic field can reach the
order of\textbf{\ $10^{19}$ }Gauss or higher in these conditions,
which is much stronger than the value $10^{16}$ Gauss in some
magnetars \cite{thom92}. The magnetic field in the interior of stars
could go up to the maximum strengths of $ 10^{18}\sim 10^{20}$ Gauss
\cite{magnetic,Dong91}. In the experiments at the LHC/CERN energy,
it is possible to produce a magnetic field of $5 \times 10^{19}$
Gauss \cite{magnetic,Voronyuk}, where all the flavors could be lying
in the lowest Landau level. Quark matter under strong magnetic field
has been studied
extensively\cite{Meneze15,Wang,Nam11,Stefan,Eduardo}. These magnetic
fields are short-lived at very high energies, but play an important
role in understanding the chiral magnetic effect, the possible
signatures of strong CP violation in experiments, and the equation
of state of compact stars\cite{Khar,Kharzeev01,Fukushima,skok09}.
Recently, the magnetic catalysis and inverse magnetic catalysis
reveal the important effect of the mangnetic effect on the chiral
symmetry restoration. The nonzero anomalous magnetic moment of
charged particle may be helpful in the realization of inverse
magnetic catalysis of the chiral transition in a strong magnetic
field, and it has been found to turn the chiral crossover into a
first-order phase transition \cite{Wen01}.
%Thermodynamic
%quantities is important in order to study the behavior of quark
%matter under the strong magnetic field.

The equation of state (EOS) is very important to characterize the
compact stars, and it has been obtained from a field theoretical
approach \cite{Broderick}. The equation of state has demonstrated
that quark stars are self-bound by the strong interaction, while
neutron stars are bound by gravity \cite{Menezes, Bandyopadhyay}. In
fact, the compressibility as an important aspect of the EoS can
reflect the relative stiffening of the equation of state of the
star. A deep understanding of the magnetic field on the equation of
state can be explored by the compressibility. Up to now, few
relevant researches have been directly done on quark matter under
strong magnetic filed. The initial work can be returned to the two
dinstict concepts of the compression modulus(incompressibility) and
the compressibility to reflect the stiffening of the equation of
state of nuclear matter and QCD matter. The compression modulus
stands for the ability to withstand compression. The larger value of
the compression modulus, the more incompressibility. However, the
compressibility is supposed as a sensitive quantity to the
fluctuation of the phase transition. Its behavior is opposite to
that of the compression modulus. Recently, much research has been
focused on nuclear matter and the quark matter under zero magnetic
field \cite{Blaizot,Blaizot01,Khoa,Kouno,Dexheimer}. The
compressibility of the rotating black hole has been investigated
\cite{Dolan}. It is found that adiabatic compressibility is maximal
in the extremal case and still less than the cold neutron star, but
it will vanish for a nonrotating black hole. The compressibility of
quark matter has been investigated in Nambu$-$Jona-Lasinio (NJL)
model and PNJL model under zero magnetic field. It is found that the
compressibility $\kappa$ is higher in NJL model than that of the
PNJL model in the hadronic phase, the behavior is just the opposite
in the partonic phase. So the core of the neutron star would be much
softer than the crust if described by PNJL model rather than NJL
model \cite{Bhattacharyya}. The compressibility are enhanced in the
chiral symmetry broken phase and are divergent at the tricritical
point in NJL model \cite{Iwasaki}. The singular behavior may be
appear in the cooling process of the QGP generated in the high
energy heavy-ion collisions.
%However, a correct running coupling that depends both on
%magnetic-field and the temperature is necessary to satisfy the
%lattice results at high magnetic field values because the
%magnetic-field and the temperature all represent energy scales in
%our consideration [arxiv:1603.03847].
% At finite temperature including the confinement mechanism,
% [this whether or not impose coupling PRD 89 116011] the Polyakov$-$Nambu$-$%
% Jona-Lasinio (PNJL)

The NJL model has proved to be very successful in the description of
the spontaneous breaking of chiral symmetry exhibited by the true
(non-perturbative) QCD vacuum, as well as many other low energy
phenomena of strong interaction. In our previous work, the strong
magnetic field effect has been investigated in the isospin asymmetry
matter and symmetric matter \cite{Yang02,Yang,Yang01}. The stronger
magnetic field can enhance the spin polarization and arrange quarks
in a uniform spin orientation \cite{Yang}.  Motivated by the work of
Miransky and Shovkovy \cite{Miransky}, it is demonstrated that the
magnetic field has key effect on the coupling constant. The inverse
magnetic catalysis can be realized by introducing the
magnetic-field-and-temperature-dependent coupling
\cite{Farias:2014eca,Ferreira:2014kpa,Andersen:2014oaa,Farias:2016gmy}.
A proper regularization scheme is required to deal with the
divergent integral in this model. In literature, the Magnetic Field
Independent Regularization (MFIR) scheme was developed in
\cite{MFIR,Frolov:2010wn} and achieved by the dimensional
regularization method in Ref. \cite{menezes09}, and extended for the
presence of color superconductivity \cite{Allen:2015paa}. The MFIR
based on the Hurwitz-Riemann zeta function was employed to study the
neutral meson in hot and magnetized quark matter
\cite{Avancini:2018svs}. In this paper, our aim is to investigate
the influence of the magnetic field on the compressibility of
two-flavor quark matter at finite densities in the framework of
SU(2)NJL model.

This work is organized as follows. In Sec. \ref{sec2}, the
thermodynamics of quark matter in a strong magnetic field is
reviewed in the NJL model. In Sec. \ref{sec:result}, the numerical
results and discussion are given with a detailed analysis of the
compressibility of quark matter as functions of the temperature, the
chemical potential and the strong magnetic fields. The last section
is a short summary.

\section{Thermodynamics of Magnetized Quark Matter in SU(2) NJL Model}
\label{sec2}

The Lagrangian density of the two-flavor NJL model in a strong
magnetic field is given as
\begin{equation}
{\mathcal{L}}_{NJL}=\bar{\psi}(i/\kern-0.5emD-m)\psi
+G[(\bar{\psi}\psi )^{2}+(\bar{\psi}i\gamma _{5}\vec{\tau}\psi
)^{2}],
\end{equation}%
where $\psi $ represents a flavor isodoublet ($u$ and $d$ quarks),
and $\vec{\tau}$ are isospin Pauli matrices. The covariant
derivative $D_{\mu }=\partial _{\mu }-iQ eA_{\mu }$ represents the
coupling of the quarks to the electromagnetic field, where
$Q=\mathrm{diag}(q_u,q_d)=\mathrm{diag}(2/3,-1/3)$ is the quark
electric charge matrix in flavor space. A sum over flavor and color
degrees of freedom is implicit. In the mean-field approximation
\cite{Ratti},
the dynamical quark mass is related to the condensation terms as%
\begin{equation}
M_{i}=m_{i}-2G\langle \bar{\psi}\psi \rangle, \label{eq:mass}
\end{equation}%
where  the current masses $m_{u}=m_{d}=m$ are used and the quark
condensates include $u$ and $d$ quark contributions as $\langle
\bar{\psi}\psi \rangle =\sum_{i=u,d}\phi _{i}$. The constituent mass
depends on both condensates. Therefore, the same mass
$M_{u}=M_{d}=M$ is available for $u$ and $d$ quarks. The
contribution from the quark flavor $i$ is
\begin{equation}
\phi _{i}=\phi _{i}^{\mathrm{vac}}+\phi _{i}^{\mathrm{mag}}+\phi _{i}^{%
\mathrm{med}}.  \label{eq:condensate}
\end{equation}%
The terms $\phi _{i}^{\mathrm{vac}}$, $\phi _{i}^{\mathrm{mag}}$, and $\phi
_{i}^{\mathrm{med}}$ representing the vacuum, magnetic field, and medium
contribution to the quark condensation are respectively \cite%
{menezes09,menezes11}
\begin{eqnarray}
\phi_i^{\mathrm{vac}} &=&-\frac{MN_{c}}{2\pi ^{2}}\left[\Lambda
\sqrt{\Lambda ^{2}+M^{2}}-M^{2}\ln (\frac{\Lambda +\sqrt{\Lambda
^{2}+M^{2}}}{M})\right],
\\
\phi _{i}^{\mathrm{mag}} &=&-\frac{M|q_{i}|eBN_{c}}{2\pi
^{2}}\left\{ \ln [\Gamma (x_{i})]-\frac{1}{2}\ln (2\pi
)+x_{i}-\frac{1}{2}(2x_{i}-1)\ln (x_{i})\right\} ,
\\
\phi _{i}^{\mathrm{med}} &=&\sum_{k_{i}=0}a_{k_{i}}\frac{M|q_{i}|eBN_{c}}{%
4\pi ^{2}}\int \frac{dp}{E_{i}^\ast}(f_i^+ + f_i^-),
\end{eqnarray}
where $a_{k_{i}}=2-\delta_{k0}$ and $k_{i}$ are respectively the
degeneracy label and the Landau quantum number. The dimensionless
quantity $x_i$ is defined as $x_i=M^{2}/(2|q_{i}|eB)$. The fermion
distribution function is
\begin{eqnarray}
f_i^\pm=\frac{1}{1+\exp[(E_i^\ast\mp\mu_i)/T]}.
\end{eqnarray}
The effective quantity $E_i^\ast=\sqrt{p^{2}+s_{i}^{2}}$ sensitively
depends on the magnetic field through $s_{i}=\sqrt{%
M^{2}+2k_{i}|q_{i}|eB}$. The chemical potential $\mu_u=\mu_d=\mu$ is
achieved for the isospin symmetry. The quark condensation is greatly
strengthened by the factor $|q_{i}eB|$ together with the dimension
reduction $D-2$ \cite{Miransky,Kojo14}. The total thermodynamic
potential density in the mean field approximation reads
\begin{equation}
\Omega =\frac{(M-m)^{2}}{4G}+\sum_{i=u,d}\Omega _{i}, \label{omega}
\end{equation}
where the first term is the interaction term. In the second term, the quantity is defined
as $\Omega _{i}=\Omega _{i}^{\mathrm{vac}}+\Omega _{i}^{%
\mathrm{mag}}+\Omega _{i}^{\mathrm{med}}$. The vacuum contribution
to the thermodynamic potential is
\begin{equation}
\Omega _{i}^{\mathrm{vac}}=\frac{N_{c}}{8\pi ^{2}}\left[ M^{4}\ln (\frac{%
\Lambda +\epsilon _{\Lambda }}{M})-\epsilon _{\Lambda }\Lambda (\Lambda
^{2}+\epsilon _{\Lambda }^{2})\right] ,
\end{equation}%
where the quantity $\epsilon _{\Lambda }$ is defined as $\epsilon
_{\Lambda }=\sqrt{\Lambda ^{2}+M^{2}}$. The ultraviolet divergence
in the vacuum part $\Omega_i^{\mathrm{vac}}$ of the thermodynamic
potential is removed by the momentum cutoff. The magnetic field and
medium contributions are respectively
\begin{equation}
\Omega _{i}^{\mathrm{mag}}=-\frac{N_{c}(|q_{i}|eB)^{2}}{2\pi
^{2}}\left[ \zeta'(-1,x_i)-\frac{1}{2}(x_{i}^{2}-x_{i})\ln (x_{i})+\frac{%
x_{i}^{2}}{4}\right],
\end{equation}%
\begin{equation}
\Omega _{i}^{\mathrm{med}}
=-T\sum_{k=0}a_{k_{i}}\frac{N_{c}|q_{i}|eB}{4\pi
^{2}}\int dp\left\{ \ln \Big[1+\exp (-\frac{E_{i}^{\ast }-\mu}{T})\Big]%
+\ln \Big[1+\exp (-\frac{E_{i}^{\ast }+\mu}{T})\Big]\right\},
\end{equation}%
where $\zeta (x,a)=\sum_{n=0}^{\infty }\frac{1}{(n+a)^{x}}$ is the
Hurwitz zeta function. From the thermodynamic potential
(\ref{omega}), one can easily obtain the quark density as
\begin{equation}
n_{i}(\mu ,B)=\sum_{k=0}a_{k_{i}}\frac{N_{c}|q_{i}|eB}{4\pi
^{2}}\int dp(f_{i}^{+}-f_{i}^{-}).
\end{equation}%
%According to the formula $S_{i}=-(\partial \Omega _{i}/\partial T)$,
%we can obtain the entropy density from the flavor $i$ contribution
%\cite{menezes11}
%\begin{equation}
%S_{i}=-\sum_{k=0}a_{k_{i}}\frac{|q_{i}|BN_{c}}{4\pi ^{2}}\int dp
%\left[f_{i}^{+}\ln (f_{i}^{+})+(1-f_{i}^{+})\ln
%(1-f_{i}^{+})+(f_{i}^{+}\leftrightarrow f_{i}^{-})\right].
%\end{equation}%
%Under strong magnetic fields, the system total pressure should be
%a sum of the matter pressure and the field pressure contribution \cite%
%{menezes09,maxwell}. But due to the requirement that the pressure
%should vanish in vacuum, the magnetic field term $B^{2}/2$ is
%automatically absent in the normalized thermodynamic quantities.
%The total pressure be written as
%\begin{equation}
%P=-\sum_i \Omega_i-\frac{(M-m_0)^2}{4G}.
%\end{equation}%

It is widely known that the strong magnetic fields can have
significant impact on the thermodynamic properties of quark matter.
Importantly, the breaking of the rotational symmetry produces the
anisotropic structure, with the parallel pressure $P^\parallel$ and
the perpendicular pressure $P^\perp$ in directions along and
transverse to the magnetic field respectively
\cite{Meneze15,Strickland:2012vu,ferr,karm,avan},
\begin{eqnarray}P^\parallel&=&-\Omega-\frac{B^2}{2}, \\
P^\perp&=&-\Omega- {\cal M}B+\frac{B^2}{2},
\end{eqnarray}
where the magnetization is ${\cal M}=-(\frac{\partial
\Omega}{\partial B})_\mu$. Consequently, the pressure anisotropy
will affect the determination of the compressibility. In the
thermodynamics, the isothermal compressibility is defined as
\begin{eqnarray}\kappa=-\frac{1}{V}(\frac{\partial V}{\partial P})_T,
\end{eqnarray} which measures how much the
volume of the system decreases with increase of pressure. The minus
is generally used for a positive quantity. Its smaller value
indicates the matter is stiffer. The compressibility of quark matter
can be derived from the potential to reflect the fluctuation of the
order parameter sensitive to the phase transition. The longitudinal
compressibility can be written in a tractable form as \cite{Iwasaki}
\begin{eqnarray} \label{eq:kapp}
\kappa^\parallel=-\frac{1}{V}(\frac{\partial V}{\partial
n_q})(\frac{\partial n_q}{\partial
P^\parallel})_T=\frac{1}{n_q^2}(\frac{\partial n_q}{\partial
\mu_q})_T,
\end{eqnarray}
where $n_q$ is the quark number density and
$\mu_q=\frac{\mu_u+\mu_d}{2}$ is the quark chemical potential. Our
compressibility reflects the thermodynamical behavior opposite to
the so-called compressibility modulus
$K=k_F^2d^2(\varepsilon/n_q)/dk_F^2$, which stands for the ability
to withstand compression. The transverse compressibility in the
direction perpendicular to the field is
\begin{eqnarray}\kappa^\perp=\frac{1}{n_q}(\frac{\partial n_q}{\partial
P^\perp})_T.
\end{eqnarray}
The compressibility could manifest the anisotropic structure due to
the breaking of the rotation symmetry. The transverse
compressibility would depend on the magnetization $\cal{M}$ in the
strong magnetic field.

\section{Numerical Results and Discussion}
 \label{sec:result}

\begin{figure}[hbt]
\begin{center}
\includegraphics[width=0.44 \textwidth]{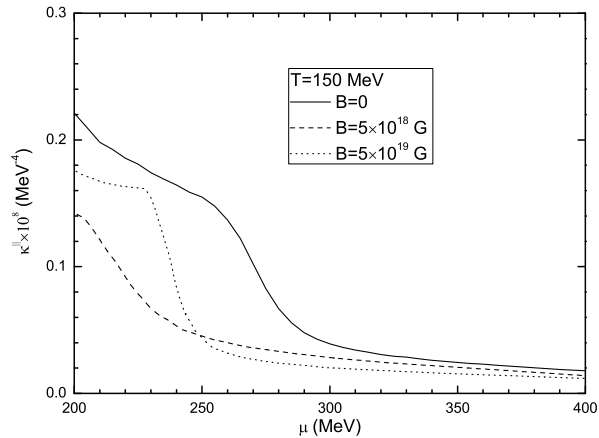}
\caption{\footnotesize The longitudinal compressibility versus the
chemical potential at different magnetic fields and a fixed
temperature.} \label{Fig1}
\end{center}
\end{figure}
In the present calculation, the following parameters are adopted:
$m_u=m_d=5.6$ MeV, $\Lambda=587.9$ MeV, and $G=2.44/\Lambda^2$. The
quark dynamical mass should be solved by the gap equation
(\ref{eq:mass}). In Fig. \ref{Fig1}, the longitudinal
compressibility are shown as the function of the chemical potential
for a fixed temperature $T=150$ MeV. The solid, dashed, and dotted
lines represent the magnetic fields $B=0, 5\times 10^{18}, 5\times
10^{19}$ Gauss, respectively. It can be clearly seen that the
magnetic field makes the compressibility become little, and
consequently results in a stiff equation of state. Especially at
smaller chemical potential, the compressibility would sensitively
depend on the magnetic field strength. At larger chemical potential
of the chiral restored phase, the magnetic effect becomes weak. This
feature can be understood by the fact that in chiral restored phase,
the quark number density is much higher and the stiffness of the
equation of state can not change evidently any more. To some extent,
it can be predicted that the magnetic field would lead to much
stiffer equation of state in neutron stars compared to zero magnetic
field condition.

\begin{figure}[hbt]
\begin{center}
\includegraphics[width=0.48 \textwidth]{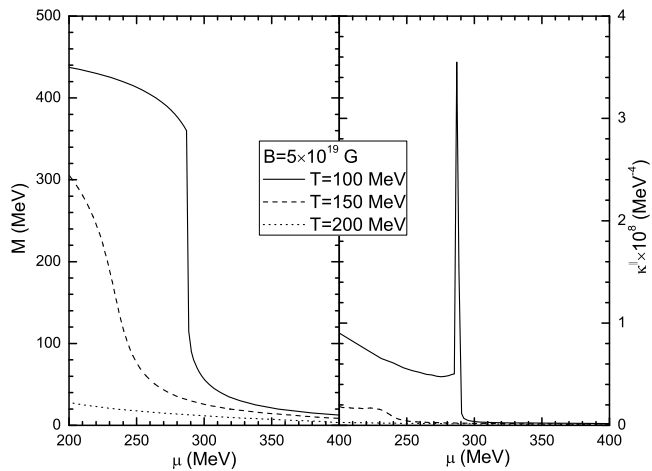}
\caption{\footnotesize The longitudinal compressibility as a
function of the chemical potential for the fixed magnetic fields and
three different temperatures.} \label{Fig2}
\end{center}
\end{figure}

In Fig. \ref{Fig2}, we show the dynamical quark mass on the left
panel and the compressibility on the right panel as the functions of
the chemical potential for a fixed magnetic field $B= 5\times
10^{19}$ Gauss. In order to investigate the thermal effect on the
compressibility, the different temperatures $T=100, 150, 200$ MeV
are marked by the solid, dashed, and dotted lines, respectively. It
is clear that the higher the temperature the less the
compressibility at a fixed magnetic field. So it is expected that
the increase of the temperature and/or the chemical potential can
make the quark matter become much stiffer. At low temperature marked
by the solid line, the compressibility changes obviously as the
chemical potential. The decreasing compressibility shows a sharp
cusp indicating the critical chemical potential of the first order
transition, which is consistent with the sharp drop of quark mass on
the left panel. However at higher temperatures, the compressibility
would remain constant and hardly change with the increasing chemical
potential any more.

\begin{figure}[ht]
\begin{center}
\includegraphics[width=0.44 \textwidth]{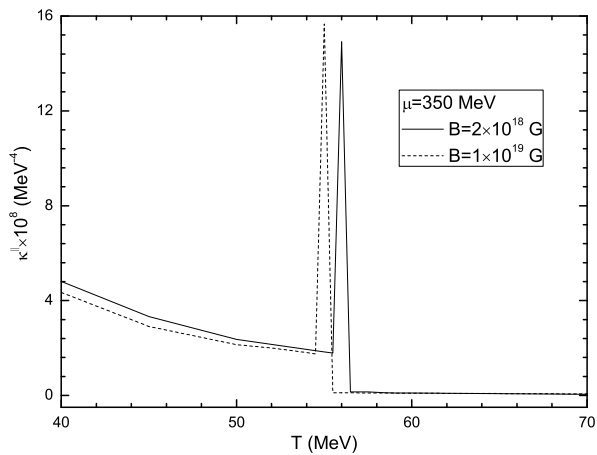}
\caption {\footnotesize The longitudinal compressibility as a
monotonous decreasing function of the temperature at the chemical
potential $\mu=350$ MeV and the magnetic fields $B=2\times10^{18}$
Gauss (solid line) and $B=1\times10^{19}$ Gauss (dashed line).}
\label{Fig3}
\end{center}
\end{figure}
To show the detailed behavior of the compressibility, we show the
compressibility as the function of the temperature at a fixed
chemical potential $\mu=350$ MeV and the two magnetic fields
$2\times 10^{18}$ and $1\times 10^{19}$ Gauss in Fig. \ref{Fig3}.
The compressibility evidently decreases as the temperature increases
and the descent behavior would slow down in the high density region.
The peak of the compressibility indicates the first-order transition
of the chiral restoration at high densities. By comparison of the
two peaks, it is indicated that for the stronger magnetic fields,
the chiral phase transition happens at lower temperature. It is
concluded that the stiffness of the equation of state can be
dominantly enhanced by the increase of the temperature at fixed
chemical potentials.

\begin{figure}[ht]
\begin{center}
\includegraphics[width=0.44 \textwidth]{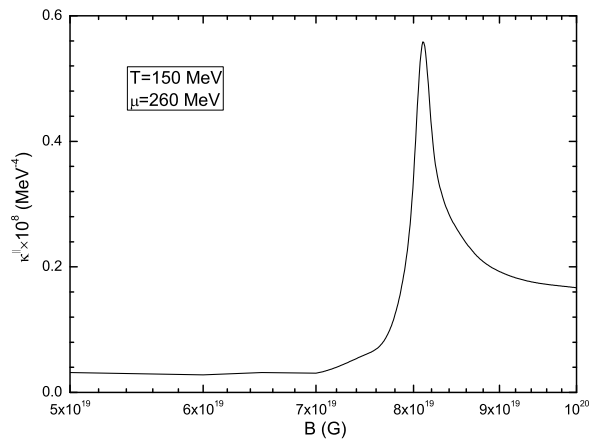}
\caption{\footnotesize Variation of longitudinal compressibility
with the magnetic field at the temperatures $T=150$ MeV and the
chemical potential $\mu=260$ MeV.} \label{Fig4}
\end{center}
\end{figure}

In Fig. \ref{Fig4}, the longitudinal compressibility versus the
magnetic field is shown at a high temperature $T=150$ MeV and the
low chemical potential $\mu=260$ MeV, the value of the
compressibility raises from a small value to a maximum and then
descends rapidly as the increasing magnetic field. The peak position
of the curve implies the chiral crossover at ($T_c$=150 MeV,
$\mu_c=260$ MeV) with the corresponding critical magnetic field
$B_c=8.1\times 10^{19}$ Gauss. From the figure we can see that a
slight perturbation of the compressibility appears due to the small
fluctuation of the order parameter at a fixed chemical potential.
The peak position could tell the precise value of the magnetic
field, at which the chiral crossover occurs at the fixed temperature
and chemical potential.

\begin{figure}[ht]
\begin{center}
\includegraphics[width=0.48 \textwidth]{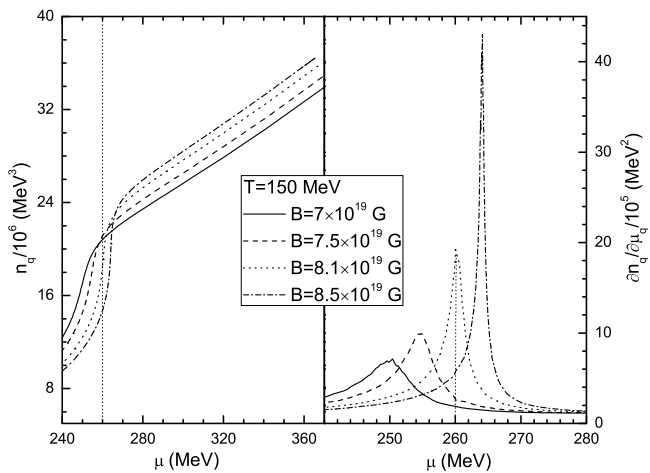}
\caption{\footnotesize The quark number density $n_q$ and the
derivatives $\partial n_q/\partial \mu_q$ versus the chemical
potential for a fixed temperatures $T=150$ MeV and four different
magnetic fields $B=7\times 10^{19}, 7.5\times 10^{19}, 8.1\times
10^{19}, 8.5\times 10^{19}$ Gauss.} \label{Fig5}
\end{center}
\end{figure}

From the definition in Eq. (\ref{eq:kapp}), it is clear that the
compressibility $\kappa$ involves the quark number susceptibility
(QNS), namely, the derivation of the quark number density with the
chemical potential ($\partial n_q/\partial \mu_q$). In literature,
the behavior of the QNS was suggested to indicate the phase
transition \cite{He:2008yr}. In Fig. \ref{Fig5}, the quark number
density and the QNS with the chemical potential are shown. For the
convenience of labeling the axis, the factors $10^{6}$ and $10^{5}$
are multiplied in the production respectively. The solid, dashed,
dotted, and dash-dotted lines stand for the magnetic fields $7\times
10^{19}$, $7.5\times 10^{19}$, $8.1\times 10^{19}$, and $8.5\times
10^{19}$ Gauss. From the numerical result, it is expected that the
quark number density increases as the chemical potential increases.
In the ascending lines on the left panel, there is a inflection
point, which can be identified by the peak of the QNS on the right
panel. The peak position of the QNS appears at ($\mu_c=260$ MeV,
$B_c=8.1\times 10^{19}$ Gauss), which exactly implies the occurrence
of the chiral crossover. It is also responsible for the peak in Fig.
\ref{Fig4}. The increasing magnetic field moves the peak position
towards the high density, which reflects the magnetic catalysis for
the pseudocritical chemical potential.

\begin{figure}[ht]
\begin{center}
\includegraphics[width=0.48 \textwidth]{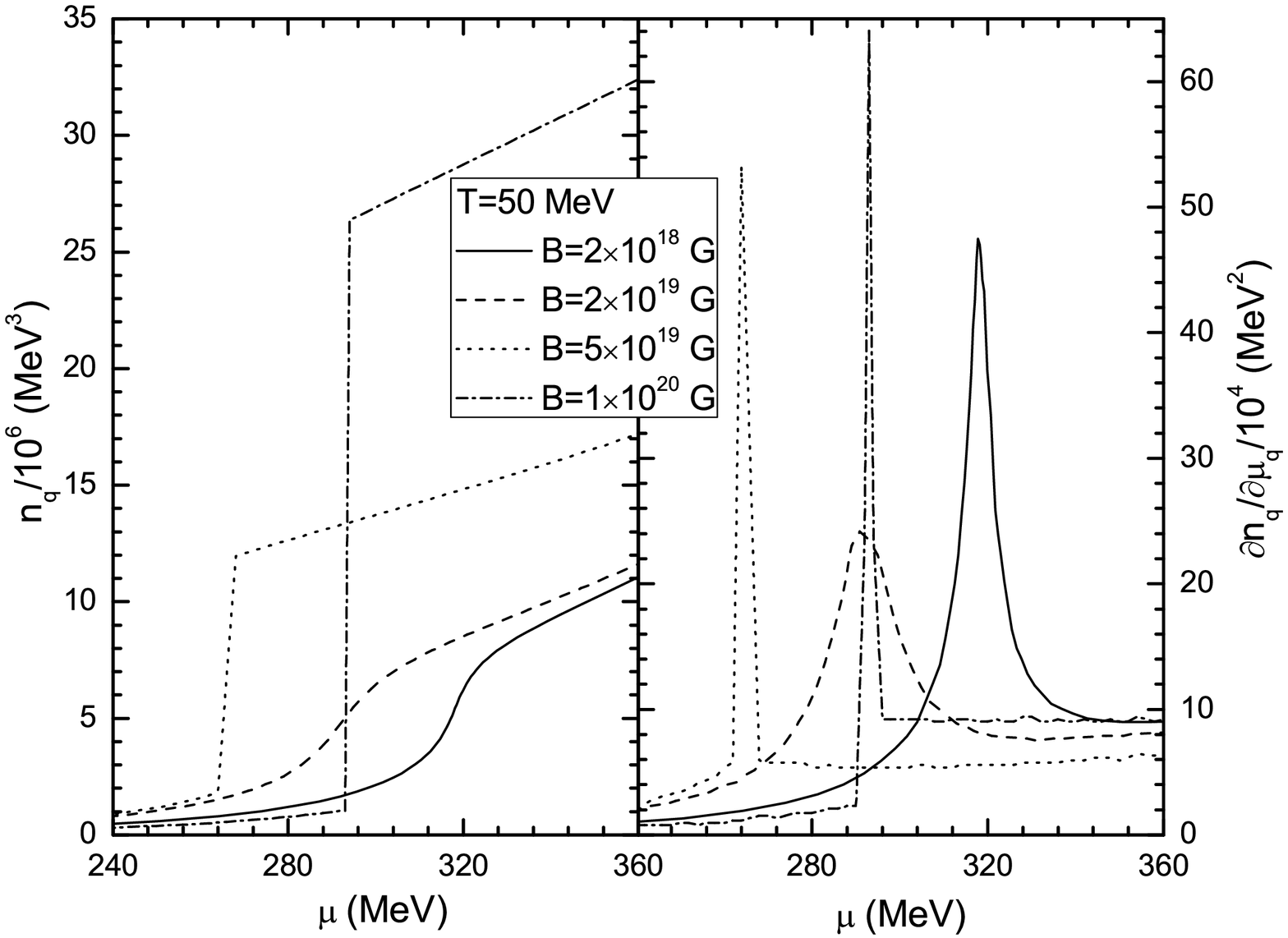}
\caption{\footnotesize The quark number density $n_q$ and the
derivatives $\partial n_q/\partial \mu_q$ versus the chemical
potential for a fixed temperatures $T=50$ MeV and four different
magnetic fields $B=2\times 10^{18}, 2\times 10^{19}, 5\times
10^{19}, 1\times 10^{20}$ Gauss.} \label{Fig6}
\end{center}
\end{figure}

\begin{figure}[ht]
\begin{center}
\includegraphics[width=0.48 \textwidth]{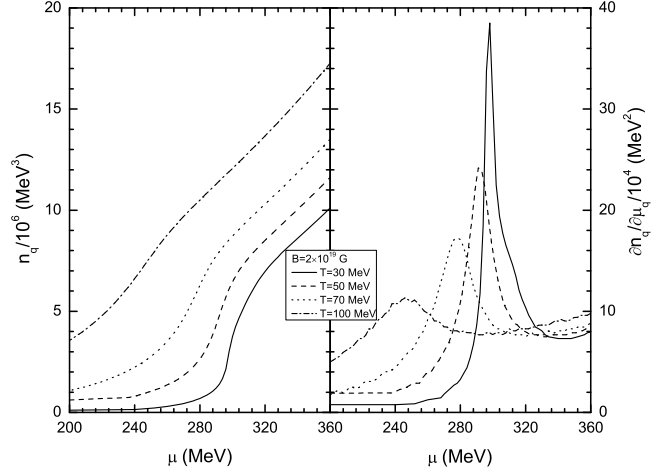}
\caption{\footnotesize The quark number density $n_q$ and the
derivatives $\partial n_q/\partial \mu_q$ versus the chemical
potential for a fixed magnetic field $B=2\times 10^{19}$ Gauss and
the four temperature $T=30, 50, 70, 100$ MeV.} \label{Fig7}
\end{center}
\end{figure}

In principle, the interaction coupling constant should be solved by
the renormalization group equation, or be phenomenologically
expressed in an effective potential dependent on environmental
variables. In the presence of a strong magnetic field, it is well
known that the interaction constant shows an obvious decreasing
behavior in addition to the enlargement of the gluon mass, which
accounts for the inverse magnetic catalysis. To obtain a reasonable
description of the temperature dependence of the interaction, we
employ the running coupling constant $G(B,T)$ from the lattice
simulations \cite{Farias:2014eca}. In Fig. \ref{Fig6} and
\ref{Fig7}, the quark density and the derivative are plotted versus
the chemical potentials. At low temperature $T=50$ MeV in Fig.
\ref{Fig6}, the peaks of the QNS indicate the first-order phase
transition at the magnetic fields $B= 5\times 10^{19}, 1\times
10^{20}$ Gauss and the crossover at the fields $B=2\times 10^{18},
2\times 10^{19}$ Gauss. The increasing magnetic fields move the
pseudocritical chemical potential to the lower value for the chiral
corssover, which indicates the inverse magnetic catalysis. On the
contrary, the magnetic catalysis on the first-order transition is
found that the the critical chemical potential is pushed to high
value by the magnetic field. In Fig. \ref{Fig7}, at the fixed
magnetic field $B=2.9\times 10^{19}$ Gauss, the thermal effect is
investigated on quark number density and the QNS. The different
temperatures $T$=30, 50, 70, and 100 MeV are marked by the solid,
the dashed, the dotted, and the dash-dotted lines, respectively. The
peaks of the QNS indicate the pseudocritical chemical potential of
the crossover is expected to be reduced by the increase of the
temperature.

\begin{figure}[ht]
\begin{center}
\includegraphics[width=0.48 \textwidth]{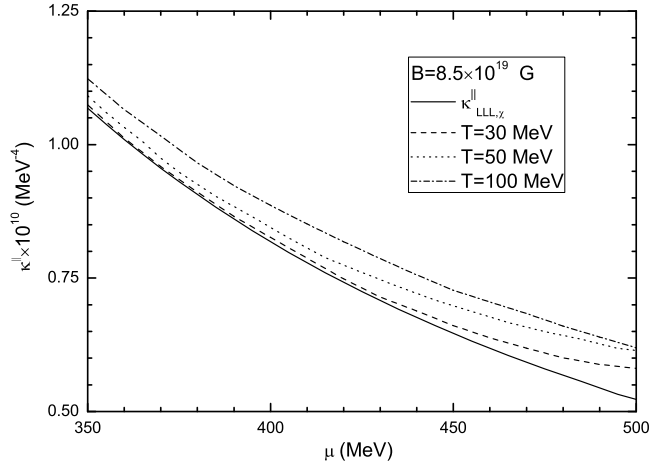}
\caption{\footnotesize The longitudinal compressibility $\kappa$
versus the chemical potential for the magnetic field $B=8.5\times
10^{19}$ Gauss and three different temperatures $T=30, 50, 100$ MeV.
The solid line indicates the compressibility
$\kappa^\parallel_{\mathrm{LLL},\chi}$ at zero temperature in the
chiral limit and the LLL approximation. } \label{Fig8}
\end{center}
\end{figure}
It is generally accepted that the strong magnetic field higher than
the order of $10^{19}$ Gauss would make the quarks occupy the lowest
landau level(LLL). Then the quark number density at $T=0$ reads,
\begin{eqnarray}n_{i}^{\mathrm{LLL}}(\mu_i ,B)=\frac{N_{c}|q_{i}|eB}{4\pi
^{2}}\sqrt{\mu_i^2-M_i^2}.
\end{eqnarray}
The corresponding derivative of the number density with respect to
the chemical potential becomes
\begin{equation}
\frac{\partial n_{i}^\mathrm{LLL}}{\partial
\mu_i}=\frac{N_{c}|q_{i}|eB}{4\pi
^{2}}\frac{\mu_i}{\sqrt{\mu_i^2-M_i^2}}.
\end{equation}
Therefore, in extreme strong magnetic fields, the isospin-symmetric
quarks in the lowest landau level contribute to the longidudinal
compressibility in a analytical form,
\begin{eqnarray}\kappa^\parallel_\mathrm{LLL}=\frac{2\pi^2}{N_{c}(|q_u|+|q_d|)eB
}\frac{\mu}{(\mu^2-M^2)^{3/2}}.
\end{eqnarray}
In the chiral limit, the dynamical quark mass vanishes and the
compressibility becomes more simply dependent on the chemical
potential,
\begin{eqnarray}\kappa^\parallel_{\mathrm{LLL}, \chi}=\frac{2\pi^2}{N_{c}(|q_u|+|q_d|)eB
\mu^2}.
\end{eqnarray}
which means that the condition of both the stronger magnetic field
and larger chemical potential would directly result in the smaller
compressibility for the bulk matter.

In Fig. \ref{Fig8}, the longitudinal compressibility
$\kappa^\parallel$ is plotted as a function of the chemical
potential at a strong magnetic field close to the order $10^{20}$
Gauss, which satisfies the lowest landau level approximation. The
degeneracy factor proportional to the field strength $|eB|$ accounts
for the larger number of quarks. The longitudinal compressibility
is a decreasing function of the chemical potential at high
densities. Moreover, it is clear that the compressibility will
gradually approach to the $\kappa^\parallel_{\mathrm{LLL},\chi}$
marked by the solid line as the temperature goes down to zero. In
the LLL approximation, the compressibility of the chiral restoration
phase at $T=0$ is inversely proportional to the chemical potential
square. Therefore, it is concluded that the increases of the
chemical potential will reduce the compressibility and lead to the
much stiffer equation of state.

\begin{figure}[ht]
\begin{center}
\includegraphics[width=0.48 \textwidth]{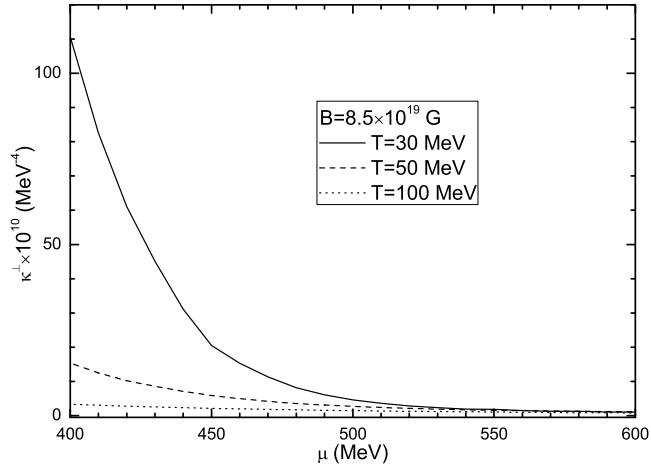}
\caption{\footnotesize The transverse compressibility $\kappa^\perp$
perpendicular to the magnetic field at $B=8.5\times 10^{19}$ Gauss
and three different temperatures $T=30, 50, 100$ MeV. } \label{Fig9}
\end{center}
\end{figure}
\begin{figure}[ht]
\begin{center}
\includegraphics[width=0.48 \textwidth]{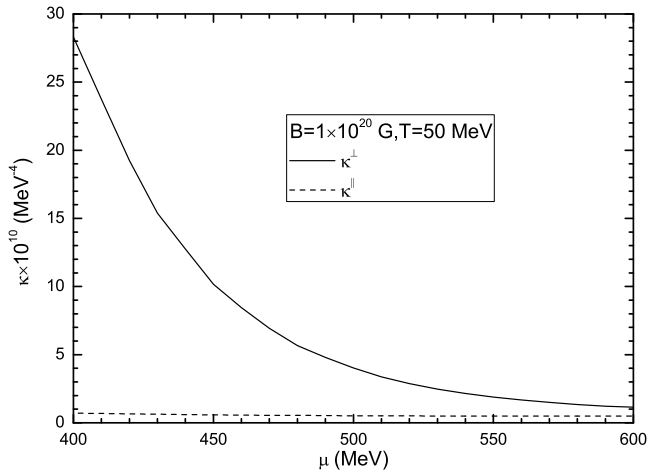}
\caption{\footnotesize The anisotropic compressibility $\kappa$
versus the chemical potential for the magnetic field $B=1.0\times
10^{20}$ Gauss and the temperature $T= 50$ MeV.  } \label{Fig10}
\end{center}
\end{figure}

The transverse compressibility as a function of the quark chemical
potential is shown in Fig. \ref{Fig9} at the magnetic field
$B=8.5\times 10^{19}$ Gauss. The descending transverse
compressibility versus the chemical potential behaves similarly to
the longitudinal compressibility. However, the anisotropic structure
induced by the external magnetic field would have important effect
on the compressibility. In Fig.~\ref{Fig10}, The longitudinal
compressibility $\kappa^\parallel$ and the transverse term
$\kappa^\perp$ are plotted at the magnetic fiend $B=1.0\times
10^{20}$GeV and the temperature $T=50$ MeV. The $\kappa^\perp$
marked by the solid line is much larger than the transverse term
marked by the dashed line. But the difference between them would
gradually become unobvious at larger chemical potentials.

\section{summary}
In this paper we have studied the compressibility of 2-flavor quark
matter in a strong magnetic field within the SU(2) NJL model. The
compressibility is defined as a variable to show how the volume
changes with the increases of the pressure. The value can reflect
the stiffness of the equation of state of quark matter. A deep
investigation on the compressibility is helpful to further
understanding the phase transition at the finite temperature, high
densities, and the strong magnetic fields. It is found that the
magnetic field can make the quark matter become stiffer by reducing
the compressibility. However, the variation of the compressibility
with the magnetic field depends on the phase that we are interested
in. The variation of the longitudinal compressibility with the
chemical potential shows a sharp cusp indicating a first-order
chiral phase transition. For the high temperature and the low
chemical potential, the peak of the longitudinal compressibility is
responsible for the critical magnetic field of the chiral crossover.

Specially, the inflection points of the quark number density and the
corresponding peak position of the quark number susceptibility
$\partial n_q/\partial \mu_q$ indicate the critical chemical
potential. By employing the magnetic-field-and-temperature-dependent
running coupling constant, firstly it is found that the increasing
magnetic field would change the chiral crossover to the first order
transition at low temperature. Secondly it is expected that the
increase of the temperature will lead to the decreasing of the
pseudocritical chemical potential for the crossover. The breaking of
the rotation leads to the anisotropic compressibility. It is found
that the transverse compressibility $\kappa^\perp$ is much larger
than the longitudinal term $\kappa^\parallel$. The difference
between the $\kappa^\perp$ and $\kappa^\parallel$ becomes unobvious
at larger chemical potential. In the very strong magnetic fields,
all quarks are lying in the lowest landau level. At zero
temperature, it has been found that the longitudinal compressibility
in the chiral restoration phase will be inversely proportional to
the magnetic field and chemical potential square, namely,
$\kappa^\parallel_{\mathrm{LLL},\chi}\propto1/(eB \mu^2)$. The
analytical relation demonstrates that the increase of the chemical
potential leads to the smaller compressibility and the stiffer
equation of state. We hope that the future investigation of the
compressibility of quark matter can provide a complementary sight to
the QCD phase at finite densities and temperatures.

\begin{acknowledgments}
The authors would like to thank support from the Scientific and
Technological Innovation Programs of Higher Education Institutions
in Shanxi under the Grant No.2020L0601 and the National Natural
Science Foundation of China (Nos. 11475110, 11875181, and 11705163).

\end{acknowledgments}

\end{document}